\input epsf

\def\figdir{.} 
\def\dancefig{cdance} 
\ifx\color+\def\dancefig{\figdir/cdance}\fi

\magnification=\magstep1
\tracingpages=1 
\baselineskip=12pt minus .1pt 

\def\TP{1} 
\def\MM{2} 
\def\B{3} 
\def\BGC{4} 
\def\MB{5} 
\def\DDH{6} 
\def\deBe{7} 
\def\deB{8} 
\def\D{9} 
\def\Fle{10} 
\def\Flo{11} 
\def\MG{12} 
\def\GJ{13} 
\def\GO{14} 
\def\G{15} 
\def\GB{16} 
\def\Gi{17} 
\def\Gii{18} 
\def\HE{19} 
\def\H{20} 
\def\HH{21} 
\def\HN{22} 
\def\J{23} 
\def\MC{24} 
\def\TTP{25} 
\def\M{26} 
\def\MW{27} 
\def\OBi{28} 
\def\OBii{29} 
\def\OBiii{30} 
\def\P{31} 
\def\RT{32} 
\def\RVZ{33} 
\def\S{34} 
\def\T{35} 
\def\W{36} 
\def\WH{37} 

\def\algoDLX{algorithm~$\rm DLX$}
\def\AlgoDLX{Algorithm~$\rm DLX$}
\def\hair{\kern.05em}
\font\csc=cmcsc10 
\font\mc=cmr9 
\font\labels=cmr9 at 9truept
\def\caption Figure #1. {{\csc Figure #1}.\enspace}
\def\newsection #1. {\medbreak\noindent{\bf#1.}\enspace}


\centerline{\bf Dancing Links}
\smallskip
\centerline{\sl Donald E. Knuth, Stanford University}
\bigskip

\noindent
My purpose is to discuss an extremely simple technique that deserves
to be better known.  Suppose~$x$ points to an element of a doubly
linked list; let $L[x]$ and $R[x]$ point to the predecessor and
successor of that element.  Then the operations
$$L\bigl[R[x]\bigr]\gets L[x], \qquad R\bigl[L[x]\bigr]\gets R[x]\eqno(1)$$
remove $x$ from the list; every programmer knows this.  But
comparatively few programmers have realized that the subsequent
operations
$$L\bigl[R[x]\bigr]\gets x, \qquad R\bigl[L[x]\bigr]\gets x\eqno(2)$$
will put $x$ back into the list again.

This fact is, of course, obvious, once it has been pointed out.  Yet I
remember feeling a definite sense of ``Aha!'' when I first realized
that (2) would work, because the values of $L[x]$ and $R[x]$ no longer
have their former semantic significance after $x$~has been removed
from its list.  Indeed, a tidy programmer might want to clean up the
data structure by setting $L[x]$ and $R[x]$ both equal to~$x$, or to
some null value, after $x$~has been deleted.  Danger sometimes lurks
when objects are allowed to point into a list from the outside; such
pointers can, for example, interfere with garbage collection.

Why, therefore, am I sufficiently fond of operation~(2) that I am
motivated to write an entire paper about it?  The element denoted
by~$x$ has been deleted from its list; why would anybody want to put
it back again?  Well, I admit that updates to a data structure are
usually intended to be permanent.  But there are also many occasions
when they are not.  For example, an interactive program may need to
revert to a former state when the user wants to undo an operation or a
sequence of operations.  Another typical application arises in {\it
backtrack programs\/} [\GB], which enumerate all solutions to a given
set of constraints.  Backtracking, also called {\it depth-first
search}, will be the focus of the present paper.

The idea of (2) was introduced in 1979 by Hitotumatu and Noshita
[\HN], who showed that it makes Dijkstra's well-known program for the
$N$~queens problem [\DDH, pages 72--82] run nearly twice as fast
without making the program significantly more complicated.

Floyd's elegant discussion of the connection between backtracking and
nondeterministic algorithms [\Flo] includes a precise method for
updating data structures before choosing between alternative lines of
computation, and for downdating the data when it is time to explore
another line.  In general, the key problem of backtrack programming
can be regarded as the task of deciding how to narrow the search and
at the same time to organize the data that controls the decisions.
Each step in the solution to a multistep problem changes the remaining
problem to be solved.

In simple situations we can simply maintain a stack that contains
snapshots of the relevant state information at all ancestors of the
current node in the search tree.  But the task of copying the entire
state at each level might take too much time.  Therefore we
often need to work with global data structures, which are modified
whenever the search enters a new level and restored when the search
returns to a previous level.

For example, Dijkstra's recursive procedure for the queens problem
kept the current state in three global Boolean arrays, representing
the columns, the diagonals, and the reverse
diagonals of a
chessboard; Hitotumatu and Noshita's program kept it in a doubly
linked list of available columns together with Boolean arrays for both
kinds of diagonals.  When Dijkstra tentatively placed a queen, he
changed one entry of each Boolean array from true to false; then he
made the entry true again when backtracking.  Hitotumatu and Noshita
used~(1) to remove a column and~(2) to restore it again; this meant
that they could find an empty column without having to search for it.
Each program strove to record the state information in such a way that
the placing and subsequent unplacing of a queen would be efficient.

The beauty of (2) is that operation (1) can be undone by knowing only
the value of~$x$.  General schemes for undoing assignments require us
to record the identity of the left-hand side together with its
previous value (see [\Flo]; see also [\TTP], pages 268--284).  But in this
case only the single quantity~$x$ is needed, and backtrack programs
often know the value of~$x$ implicitly as a byproduct of their normal
operation.

We can apply (1) and (2) repeatedly in complex data structures that
involve large numbers of interacting doubly linked lists.  The program
logic that traverses those lists and decides what elements should be
deleted can often be run in reverse, thereby deciding what elements
should be undeleted.  And undeletion restores links that allow us to
continue running the program logic backwards until we're ready to go
forward again.

This process causes the pointer variables inside the
global data structure to execute an exquisitely choreographed dance;
hence I like to call~(1) and~(2) the technique of {\it dancing
links}.

\newsection The exact cover problem.
One way to illustrate the power of dancing links is to consider a
general problem that can be described abstractly as follows: Given a
matrix of 0s and 1s, does it have a set of rows containing exactly one
1 in each column?

For example, the matrix
$$\pmatrix{0&0&1&0&1&1&0\cr 1&0&0&1&0&0&1\cr 0&1&1&0&0&1&0\cr
1&0&0&1&0&0&0\cr 0&1&0&0&0&0&1\cr 0&0&0&1&1&0&1\cr}\eqno(3)$$
has such a set (rows 1, 4, and 5).  We can think of the columns as
elements of a universe, and the rows as subsets of the universe; then
the problem is to cover the universe with disjoint subsets.  Or we can
think of the rows as elements of a universe, and the columns as
subsets of that universe; then the problem is to find a collection of
elements that intersect each subset in exactly one point.  Either way,
it's a potentially tough problem, well known to be NP-complete even
when each row contains exactly three 1s [\GJ, page 221].  And it is a
natural candidate for backtracking.

Dana Scott conducted one of the first experiments on backtrack
programming in 1958, when he was a graduate student at Princeton
University [\S].  His program, written for the IAS ``{\mc MANIAC}'' computer
with the help of Hale~F. Trotter, produced the first listing of all
ways to place the 12~pentominoes into a chessboard leaving the center
four squares vacant.  For example, one of the 65~solutions is shown in
Figure~1.  (Pentominoes are the case $n=5$ of $n$-ominoes, which are
connected $n$-square subsets of an infinite board; see~[\G]. Scott was
probably inspired by Golomb's paper~[\GO] and some extensions
reported by Martin Gardner~[\MG].)

\midinsert
\line{\hfil\epsfxsize=.3\hsize \epsfbox{\dancefig.1}\hfil
              \caption Figure 1. Scott's pentomino problem.\hfil}
\endinsert

This problem is a special case of the exact cover problem.  Imagine a
matrix that has 72 columns, one for each of the 12~pentominoes and one
for each of the 60~cells of the chessboard-minus-its-center.
Construct all possible rows representing a way to place a pentomino on
the board; each row contains a~1 in the column identifying the piece,
and five~1s in the columns identifying its positions.  (There are
exactly 1568 such rows.)  We can name the first twelve columns ${\rm
F\,I\,L\,P\,N\,T\,U\,V\,W\,X\,Y\,Z}$, following Golomb's recommended
names for the pentominoes [\G, page 7], and we can use two digits $ij$
to name the column corresponding to rank~$i$ and file~$j$ of the
board; each row is conveniently represented by giving the names of the
columns where 1s appear.  For example, Figure~1 is the exact cover
corresponding to the twelve rows
$$\matrix{{\rm I}&11&12&13&14&15\cr
          {\rm N}&16&26&27&37&47\cr
          {\rm L}&17&18&28&38&48\cr
          {\rm U}&21&22&31&41&42\cr
          {\rm X}&23&32&33&34&43\cr
          {\rm W}&24&25&35&36&46\cr
          {\rm P}&51&52&53&62&63\cr
          {\rm F}&56&64&65&66&75\cr
          {\rm Z}&57&58&67&76&77\cr
          {\rm T}&61&71&72&73&81\cr
          {\rm V}&68&78&86&87&88\cr
          {\rm Y}&74&82&83&84&85\rlap{\thinspace.}\cr}$$

\newsection Solving an exact cover problem.
The following nondeterministic algorithm, which I will call algorithm~X
for lack of a better name, finds all solutions to the
exact cover problem defined by any given matrix~$A$ of 0s and 1s.
Algorithm~X is simply a statement of the
obvious trial-and-error approach.  (Indeed, I can't think of any other
reasonable way to do the job, in general.)

\bigbreak
\begingroup\obeylines
If $A$ is empty, the problem is solved; terminate successfully.
Otherwise choose a column, $c$ (deterministically).
Choose a row, $r$, such that $A[r,c]=1$ (nondeterministically).
Include $r$ in the partial solution.
For each $j$ such that $A[r,j]=1$,
\qquad delete column $j$ from matrix~$A$;
\qquad for each $i$ such that $A[i,j]=1$,
\qquad\qquad delete row $i$ from matrix~$A$.
Repeat this algorithm recursively on the reduced matrix~$A$.
\endgroup
\bigbreak\noindent
The nondeterministic choice of $r$ means that the algorithm
essentially clones itself into independent subalgorithms; each
subalgorithm inherits the current matrix~$A$, but reduces it with
respect to a different row~$r$.  If column~$c$ is entirely zero, there
are no subalgorithms and the process terminates unsuccessfully.

The
subalgorithms form a {\it search tree\/} in a natural way, with the
original problem at the root and with level~$k$ containing each subalgorithm
that corresponds to~$k$ chosen rows.  Backtracking is the
process of traversing the tree in preorder, ``depth first.''

Any systematic rule for choosing column~$c$ in this procedure will
find all solutions, but some rules work much better than others.  For
example, Scott~[\S] said that his initial inclination was to place the first
pentomino first, then the second pentomino, and so~on; this would
correspond to choosing column~${\rm F}$ first, then column~${\rm I}$,
etc., in the corresponding exact cover problem.  But he soon realized
that such an approach would be hopelessly slow: There are 192~ways to
place the~~${\rm F}$, and for each of these there are approximately
34~ways to place the~${\rm I}$.  The Monte Carlo estimation procedure
described in [\MC] suggests that the search tree for this scheme has
roughly $2\times10^{12}$ nodes!  By contrast, the alternative of
choosing column~11 first (the column corresponding to rank~1 and
file~1 of the board), and in general choosing the
lexicographically first uncovered column, leads to a search tree with
9{,}015{,}751 nodes.

Even better is the strategy that Scott finally adopted~[\S]:
He realized that piece~${\rm X}$ has only 3 essentially
different positions, namely centered at 23, 24, and 33.  Furthermore,
if the $X$ is at 33, we can assume that the ${\rm P}$~pentomino is not
``turned over,'' so that it takes only four of its eight orientations.
Then we get each of the 65~essentially different solutions exactly
once, and the full set of $8\times65=520$ solutions is easily obtained
by rotation and reflection.  These constraints on~${\rm X}$ and~${\rm
P}$ lead to three independent problems, with

$$\eqalign{103{,}005 {\rm \ nodes\ and\ } 19 {\rm \ solutions} 
\quad &({\rm X\ at\ }23);\cr
           106{,}232 {\rm \ nodes\ and\ } 20 {\rm \ solutions} 
\quad &({\rm X\ at\ }24);\cr
           126{,}636 {\rm \ nodes\ and\ } 26 {\rm \ solutions} 
\quad &({\rm X\ at\ }33, {\rm \ P\ not\ flipped),}}$$
when columns are chosen lexicographically.

Golomb and Baumert [\GB] suggested choosing, at each stage of a
backtrack procedure, a subproblem that leads to the fewest branches,
whenever this can be done efficiently.  In the case of an exact cover
problem, this means that we want to choose at each stage a column with
fewest~1s in the current matrix~$A$.  Fortunately we will see
that the technique of dancing links allows us to do this quite nicely;
the search trees for Scott's pentomino problem then have only
$$\eqalign{10{,}421 {\rm \ nodes} \quad &({\rm X\ at\ }23);\cr
           12{,}900 {\rm \ nodes} \quad &({\rm X\ at\ }24);\cr
           14{,}045 {\rm \ nodes} \quad &({\rm X\ at\ }33, {\rm \ P\ not\ flipped),}}$$
respectively.

\newsection The dance steps.
One good way to implement algorithm~$X$ is to represent each 1 in the
matrix~$A$ as a {\it data object\/}~$x$ with five fields $L[x],
R[x], U[x], D[x], C[x]$.  Rows of the matrix are doubly linked as
circular lists via the $L$ and $R$~fields (``left'' and ``right'');
columns are doubly linked as circular lists via the $U$ and $D$~fields
(``up'' and ``down'').  Each column list also includes a special data
object called its {\it list header}.

The list headers are part of a larger object called a {\it column
object}.  Each column object~$y$ contains the fields $L[y], R[y],
U[y], D[y]$, and $C[y]$ of a data object and two additional fields,
$S[y]$ (``size'') and $N[y]$ (``name''); the size is the number of 1s
in the column, and the name is a symbolic identifier for printing the
answers.  The $C$~field of each object points to the column object at
the head of the relevant column.

The $L$ and $R$ fields of the list headers link together all columns
that still need to be covered.  This circular list also includes a
special column object called the {\it root}, $h$, which serves as a
master header for all the active headers.  The fields $U[h], D[h],
C[h], S[h]$, and $N[h]$ are not used.

For example, the 0\hair-1 matrix of (3) would be represented by the
objects shown in Figure~2, if we name the columns {\rm A, B, C, D, E,
F, and G}.  (This diagram ``wraps around'' toroidally at the top,
bottom, left, and right.  The $C$~links are not shown because they
would clutter up the picture; each $C$~field points to the topmost
element in its column.)

\topinsert
\epsfxsize=.55\hsize
\centerline{\epsfbox{\dancefig.2}}
\medskip
\centerline{\caption Figure 2. Four-way-linked representation of the
  exact cover problem (3).}
\endinsert

\def\\#1{\leavevmode\hbox{\it#1\kern.05em}}
Our nondeterministic algorithm to find all exact covers can now be
cast in the following explicit, deterministic form as a recursive
procedure \\{search}$(k)$, which is invoked initially with $k=0$:

\medskip
\begingroup\obeylines
If $R[h]=h$, print the current solution (see below) and return.
Otherwise choose a column object $c$ (see below).
Cover column $c$ (see below).
For each $r\gets D[c]$, $D\bigl[D[c]\bigr]$, \dots, while $r\ne c$,
\qquad set $O_k\gets r$;
\qquad for each $j\gets R[r]$, $R\bigl[R[r]\bigr]$, \dots, while $j\ne r$,
\qquad\qquad cover column $j$ (see below);
\qquad \\{search}$(k+1)$;
\qquad set $r\gets O_k$ and $c\gets C[r]$;
\qquad for each $j\gets L[r]$, $L\bigl[L[r]\bigr]$, \dots, while $j\ne r$,
\qquad\qquad uncover column $j$ (see below).
Uncover column $c$ (see below) and return.
\endgroup
\medbreak
\noindent
The operation of printing the current solution is easy: We successively print
the rows containing $O_0$, $O_1$, \dots,~$O_{k-1}$, where the row
containing data object~$O$ is printed by printing $N\bigl[C[O]\bigr]$,
$N\bigl[C[R[O]]\bigr]$, $N\bigl[C[R[R[O]]]\bigr]$, etc.

To choose a column object $c$, we could simply set $c\gets R[h]$;
this is the leftmost uncovered column.
Or if we want to minimize the branching factor, we could set
$s\gets\infty$ and then
$$\vbox{\halign{#\hfil\cr
for each $j\gets R[h]$, $R\bigl[R[h]\bigr]$, \dots,
          while $j\ne h$,\cr
\qquad if $S[j]<s$ \quad set $c\gets j$ and $s\gets S[j]$.\cr}}$$
Then $c$ is a column with the smallest number of 1s.  (The $S$ fields
are not needed unless we want to minimize branching in this way.)

The operation of covering column~$c$ is more interesting: It
removes~$c$ from the header list and removes all rows in $c$'s~own
list from the other column lists they are in.

\medskip
\begingroup\obeylines
Set $L\bigl[R[c]\bigr]\gets L[c]$ and $R\bigl[L[c]\bigr]\gets R[c]$.
For each $i\gets D[c]$, $D\bigl[D[c]\bigr]$, \dots, while $i\ne c$,
\qquad for each $j\gets R[i]$, $R\bigl[R[i]\bigr]$, \dots, while $j\ne i$,
\qquad\qquad set $U\bigl[D[j]\bigr]\gets U[j]$, \ $D\bigl[U[j]\bigr]\gets D[j]$,
\qquad\qquad and set $S\bigl[C[j]\bigr]\gets S\bigl[C[j]\bigr]-1$.
\endgroup
\medskip
\noindent
Operation (1), which I mentioned at the outset of this paper, is used
here to remove objects in both the horizontal and vertical directions.

Finally, we get to the point of this whole algorithm, the operation of
{\it uncovering\/} a~given column~$c$.  Here is where the links do their
dance:

\medskip
\begingroup\obeylines
For each $i=U[c]$, $U\bigl[U[c]\bigr]$, \dots, while $i\ne c$,
\qquad for each $j\gets L[i]$, $L\bigl[L[i]\bigr]$, \dots, while $j\ne i$,
\qquad\qquad set $S\bigl[C[j]\bigr]\gets S\bigl[[j]\bigr] + 1$,
\qquad\qquad and set $U\bigl[D[j]\bigr]\gets j$, \ $D\bigl[U[j]\bigr]\gets j$.
Set $L\bigl[R[c]\bigr]\gets c$ and $R\bigl[L[c]\bigr]\gets c$.
\endgroup
\medbreak
\noindent
Notice that uncovering takes place in precisely the reverse order of
the covering operation, using the fact that (2)~undoes (1).  (Actually
we need not adhere so strictly to the principle of ``last done, first undone''
in this case, since $j$~could run through row~$i$ in any order.
But we must be careful
to unremove the rows from bottom to top, because we removed them from
top to bottom.  Similarly, it is important to uncover the columns of
row~$r$ from right to left, because we covered them from left to
right.)

\midinsert
\epsfxsize=.55\hsize
\centerline{\epsfbox{\dancefig.3}}
\medskip
\centerline{\caption Figure 3. The links after column~A in Figure~2 has been covered.}
\endinsert

Consider, for example, what happens when \\{search}$(0)$ is applied to
the data of~(3) as represented by Figure~2.  Column~${\rm A}$ is
covered by removing both of its rows from their other columns; the
structure now takes the form of Figure~3. Notice the asymmetry of the
links that now appear in column~D: The upper element was deleted
first, so it still points to its original neighbors, but the other
deleted element points upward to the column header.

Continuing \\{search}(0), when $r$~points to
the ${\rm A}$~element of row~${\rm (A,D,G)}$, we also cover
columns~$\rm D$ and $\rm G$.  Figure~4 shows the status as we enter
\\{search}(1); this data structure represents the reduced matrix
$$\bordermatrix{&{\rm B}&{\rm C}&{\rm E}&{\rm F}\cr
  &0&1&1&1\cr
  &1&1&0&1}.\eqno(4)$$

\topinsert
\epsfxsize=.55\hsize
\centerline{\epsfbox{\dancefig.4}}
\medskip
\centerline{\caption Figure 4. The links after columns D and~G in Figure~3
   have been covered.}
\endinsert

Now \\{search}(1) will cover column~${\rm B}$, and there will be no 1s
left in column~${\rm E}$.  So \\{search}(2) will find nothing.  Then
\\{search}(1) will return, having found no solutions, and the state of
Figure~4 will be restored.  The outer level routine, \\{search}(0),
will proceed to convert Figure~4 back to Figure~3, and it will
advance~$r$ to the ${\rm A}$~element of row~(${\rm A,D}$).

Soon the solution will be found.  It will be printed as 
$$\matrix{{\rm A}&{\rm D}\cr
          {\rm B}&{\rm G}\cr
          {\rm C}&{\rm E}&{\rm F}}$$
if the $S$~fields are ignored in the choice of~$c$, or as
$$\matrix{{\rm A}&{\rm D}\cr
          {\rm E}&{\rm F}&{\rm C}\cr
          {\rm B}&{\rm G}}$$
if the shortest column is chosen at each step.  (The first item
printed in each row list is the name of the column on which branching
was done.)  Readers who play through the action of this algorithm on
some examples will understand why I chose the title of this paper.

\newsection Efficiency considerations.
When algorithm~$\rm X$ is implemented in terms of dancing links, let's
call it \algoDLX.  The running time of \algoDLX\ is essentially
proportional to the number of times it applies operation~(1) to remove
an object from a list; this is also the number of times it applies
operation~(2) to unremove an object.  Let's say that this quantity is the
number of {\it updates}.  A total of 28~updates are performed during
the solution of~(3) if we repeatedly choose the shortest
column: 10~updates are made on level~0, 14~on level~1, and 4~on
level~2.  Alternatively, if we ignore the $S$~heuristic, the algorithm
makes 16~updates on level~1 and 7~updates on level~2, for a total of~33.  But
in the latter case each update will go noticeably faster, since the
statements $S\bigl[C[j]\bigr]\gets S\bigl[C[j]\bigr]\pm1$ can be
omitted; hence the overall running time will probably be less.  Of
course we need to study larger examples before drawing any general
conclusions about the desirability of the $S$~heuristic.

\topinsert
\epsfysize=3.6truein
\epsfxsize=\hsize
\scrollmode
\epsfbox{\figdir/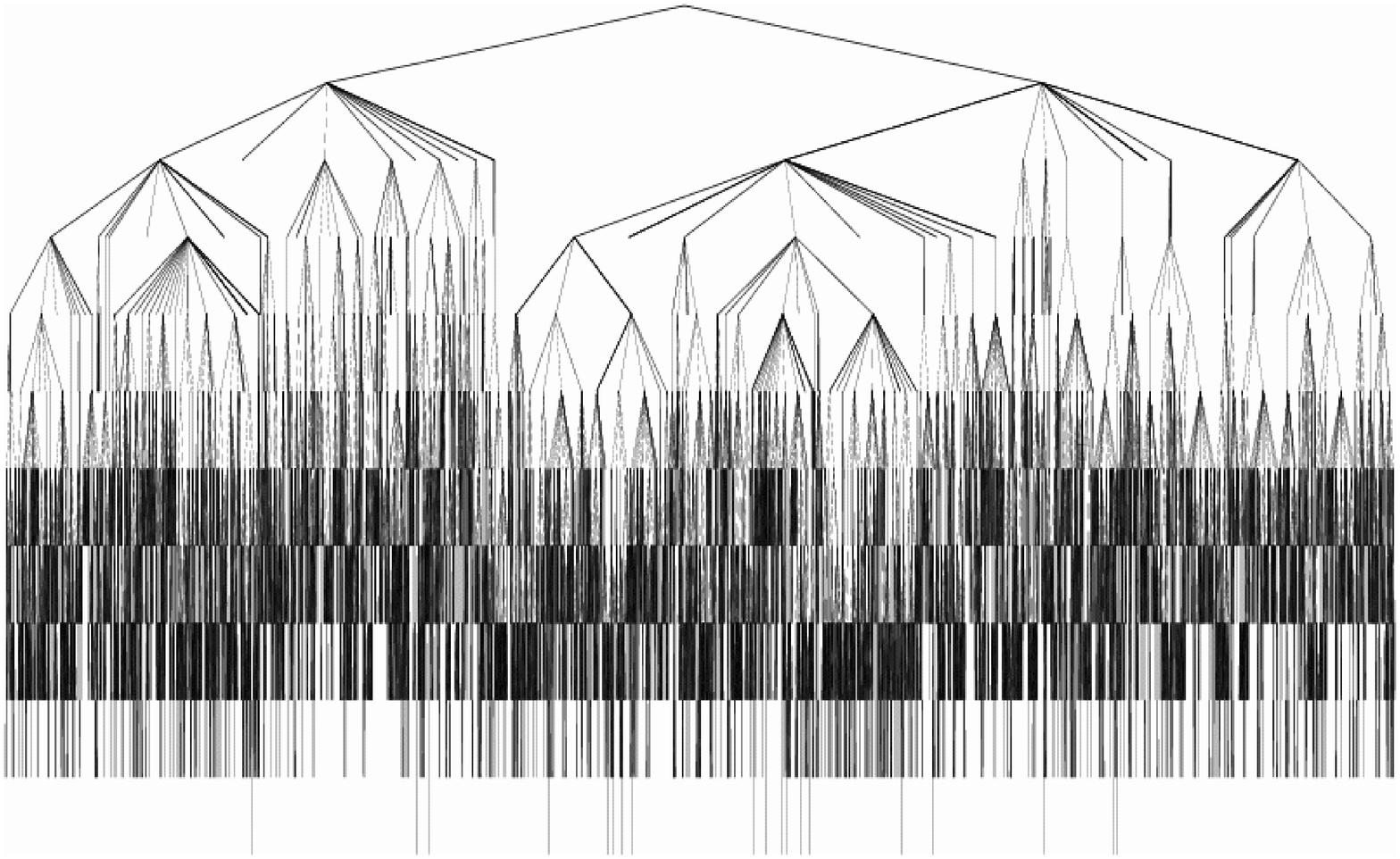} 
\errorstopmode
\medskip
\centerline{\caption Figure 5. The search tree for one case of
   Scott's pentomino problem.}
\endinsert

A backtrack program usually spends most of its time on only a few
levels of the search tree (see [\MC]).  For example, Figure~5 shows
the search tree for the case $\rm X=23$ of Dana Scott's pentomino problem
using the $S$~heuristic;
it has the following profile:
$$\vbox{\halign{\hfil#\quad&\hfil#\quad&(\hfil#\%)\quad&\hfil#\quad&(\hfil#\%)\quad&\hfil#\cr
Level&\multispan2\hfil Nodes\hfil&\multispan2\hfil Updates\hfil&Updates per \rlap{node}\cr
\noalign{\vskip2pt}
 0&    1&  0&   2,031&  0& 2031.0\cr
 1&    2&  0&   1,676&  0&  838.0\cr
 2&   22&  0&  28,492&  1& 1295.1\cr
 3&   77&  1&  77,687&  2& 1008.9\cr
 4&  219&  2& 152,957&  4&  698.4\cr
 5&  518&  5& 367,939& 10&  710.3\cr
 6& 1,395& 13& 853,788& 24&  612.0\cr
 7& 2,483& 24& 941,265& 26&  379.1\cr
 8& 2,574& 25& 740,523& 20&  287.7\cr
 9& 2,475& 24& 418,334& 12&  169.0\cr
10&  636&  6&  32,205&  1&   50.6\cr
11&   19&  0&    826&  0&   43.5\cr
\noalign{\vskip2pt}
Total&10,421&\omit\hidewidth(100\%)\quad&\hfil 3,617,723&\omit\hidewidth(100\%)\quad&347.2\cr}}$$
(The number of updates shown for level~$k$ is the number of times an
element was removed from a doubly linked list during the calculations
between levels~$k-1$ and~$k$.  The 2,031~updates on level~0 correspond
to removing column~$\rm X$ from the header list and then removing
$2030/5=406$~rows from their other columns; these are the rows that
\vadjust{\goodbreak}%
overlap with the placement of~$\rm X$ at~23.  A slight optimization was
made when tabulating this data: Column~$c$ was not covered and
uncovered in trivial cases when it contained no rows.)  Notice that
more than half of the nodes lie on levels~$\ge8$, but more than half
of the updates occur on the way to level~7.  Extra work on the lower
levels has reduced the need for hard work at the higher levels.

The corresponding statistics look like this when the same problem is
run {\it without\/} the ordering heuristic based on $S$~fields:
$$\vbox{\halign{\hfil#\quad&\hfil#\quad&(\hfil#\%)\quad&\hfil#\quad&(\hfil#\%)\quad&\hfil#\cr
Level&\multispan2\hfil Nodes\hfil&\multispan2\hfil Updates\hfil&Updates per \rlap{node}\cr
\noalign{\vskip2pt}
0& 1& 0& 2{,}031& 0& 2031.0\cr
1& 6& 0& 5{,}606& 0& 934.3\cr
2& 24& 0& 30{,}111& 0& 1254.6\cr
3& 256& 0& 249{,}904& 1& 976.2\cr
4& 581& 1& 432{,}471& 2& 744.4\cr
5& 1{,}533& 1& 1{,}256{,}556& 7& 819.7\cr
6& 3{,}422& 3& 2{,}290{,}338& 13& 669.3\cr
7& 10{,}381& 10& 4{,}442{,}572& 25& 428.0\cr
8& 26{,}238& 25& 5{,}804{,}161& 33& 221.2\cr
9& 46{,}609& 45& 3{,}006{,}418& 17& 64.5\cr
10& 13{,}935& 14& 284{,}459& 2& 20.4\cr
11& 19& 0& 14{,}125& 0& 743.4\cr
\noalign{\vskip2pt}
Total&103{,}005&\omit\hidewidth(100\%)\quad&17{,}818{,}752&\omit\hidewidth(100\%)\quad&173.0\cr}}$$
Each update involves about 14 memory accesses when the $S$~heuristic
is used, and about 8 accesses when $S$~is ignored.  Thus the
$S$~heuristic multiplies the total number of memory accesses by a factor
of approximately $(14\times3{,}617{,}723)/(8\times17{,}818{,}752)\approx36\%$ in this
example.  The heuristic is even more effective in larger problems,
because it tends to reduce the total number of nodes by a factor that
is exponential in the number of levels while the cost of applying it
grows only linearly.

Assuming that the $S$~heuristic is good in large trees but not so good
in small ones, I tried a hybrid scheme that uses~$S$ at low levels but
not at high levels.  This experiment was, however, unsuccessful.  If,
for example, $S$~was ignored after level~7, the statistics for
levels~8--11 were as follows:
$$\vbox{\halign{\hfil#\qquad&\hfil#\quad&\hfil#\cr
Level&Nodes&Updates\cr
\noalign{\vskip2pt}
8&18{,}300&5{,}672{,}258\cr
9&28{,}624&2{,}654{,}310\cr
10&9{,}989&213{,}944\cr
11&19&10{,}179\cr}}$$
And if the change was applied after level 8, the statistics were
$$\postdisplaypenalty=10000
\vbox{\halign{\hfil#\qquad&\hfil#\quad&\hfil#\cr
Level&Nodes&Updates\cr
\noalign{\vskip2pt}
9&11,562&1,495,054\cr
10&6,113&148,162\cr
11&19&6,303\cr}}$$
Therefore I decided to retain the $S$~heuristic at all levels of \algoDLX.

My trusty old {\mc SPARC}station 2 computer, vintage 1992, is able to
perform approximately 0.39~mega-updates per second when working on
large problems and maintaining the $S$~fields.  The 120~MHz Pentium~I
computer that Stanford computer science faculty were given
in 1996 did 1.21~mega-updates per second, and
my new 500~MHz Pentium~III does 5.94.  Thus the running time decreases
as technology advances; but it remains essentially proportional to the
number of updates, which is the number of times the links do their
dance.  Therefore I prefer to measure the performance of \algoDLX\ by
counting the number of updates, not by counting the number of elapsed seconds.

Scott [\S] was pleased to discover that his program for the {\mc MANIAC}
solved the pentomino problem in about 3.5~hours.  The {\mc MANIAC} executed
approximately 4000~instructions per second, so this represented
roughly 50~million instructions.  He and H.~F. Trotter found a nice way
to use the ``bitwise-and'' instructions of the
{\mc MANIAC}, which had 40-bit registers.  Their code, which executed about
$50{,}000{,}000/(103{,}005+106{,}232+154{,}921)\approx 140$ instructions per node of
the search tree, was quite efficient in spite of the fact that they
had to deal with about ten times as many nodes as would be produced by
the ordering heuristic.  Indeed, the linked-list approach of \algoDLX\
performs a total of $3{,}617{,}723 + 4{,}547{,}186 + 5{,}526{,}988 = 13{,}691{,}897$
updates, or about 192~million memory accesses; and it would never fit
in the 5120-byte memory of the {\mc MANIAC}!  From this standpoint the
technique of dancing links is actually a step backward from Scott's
40-year-old method, although of course that method works only for
very special types of exact cover problems in which simple geometric
structure can be exploited.

The task of finding all ways to pack the set of pentominoes into a
$6\times 10$~rectangle is more difficult than Scott's
$8\times8-2\times2$ problem, because the backtrack tree for the
$6\times 10$ problem is larger and there are 2339 essentially different
solutions~[\HH].  In this case we limit the $\rm X$~pentomino to the upper
left quarter of the board; our linked-memory algorithm generates 902{,}631
nodes and 309{,}134{,}131 updates (or 28{,}320{,}810 nodes and
4{,}107{,}105{,}935 updates without the $S$~heuristic).  This solves
the problem in less than a minute on a Pentium~III; however, again I
should point out that the special characteristics of pentominoes allow
a faster approach.

John G.\ Fletcher needed only ten minutes to solve
the $6\times10$ problem on an IBM~7094 in 1965, using a highly optimized program
that had 765~instructions in its inner loop [\Fle].  The 7094~had a clock
rate of 0.7~MHz, and it could access two 36-bit words in a single
clock cycle.  Fletcher's program required only about $600\times
700{,}000/28{,}320{,}810\approx 15$ clock cycles per node of the
search tree; so it
was superior to the bitwise method of Scott and Trotter, and it
remains the fastest algorithm known for problems that involve placing
the twelve pentominoes. (N.~G. de~Bruijn discovered an almost identical
method independently; see~[\deBe].)

With a few extensions to the 0\hair-1 matrix for Dana Scott's problem, we
can solve the more general problem of covering a chessboard with twelve
pentominoes and one square tetromino, without insisting that the
tetromino occupy the center.  This is essentially the classic problem
of Dudeney, who invented pentominoes in 1907 [\D].  The total number
of such chessboard dissections has apparently never appeared in the
literature; \algoDLX\ needs 1{,}526{,}279{,}783 updates to determine
that it is exactly 16{,}146.

Many people have written about polyomino problems, including
distinguished mathematicians such as Golomb [\G],
de~Bruijn [\deBe,\thinspace\deB],
Berlekamp, Conway and Guy [\BGC].  Their arguments for placing the
pieces are sometimes based on enumerating the number of ways a certain
cell on the board can be filled, sometimes on the number of ways a
certain piece can be placed.  But as far as I know, nobody has
previously pointed out that such problems are actually exact cover
problems, in which there is perfect symmetry between cells and pieces.
\AlgoDLX\ will branch on the ways to fill a cell if some cell is
difficult to fill, or on the ways to place a piece if some piece is
difficult to place.  It knows no difference, because pieces and cells
are simply columns of the given input matrix.

\topinsert
\vskip-6pt
\def\\#1#2#3#4{
\vbox{
 \hsize=.37\hsize 
 \epsfxsize=\hsize
 \epsfbox{\dancefig.6#1}
 \smallskip
 \centerline{\labels #2 solutions, #3 nodes, #4 updates}}}

\line{\hfil\\1{92}{14,352,556}{1,764,631,796}\hfil\qquad
      \\2{100}{10,258,180}{1,318,478,396}\hfil}
\medskip
\line{\hfil\\3{20}{6,375,335}{806,699,079}\hfil\qquad
      \\4{0}{1,234,485}{162,017,125}\hfil}
\medskip
\centerline{\caption Figure 6. Packing 45 Y pentominoes into a square.}
\vskip-8pt
\endinsert

\AlgoDLX\ begins to outperform other pentomino-placing procedures in
problems where the search tree has many levels.  For example, let's
consider the problem of packing 45~$\rm Y$~pentominoes into a $15\times15$
square.  Jenifer Haselgrove studied this with the help of a machine
called the ICS Multum---which qualified as a ``fast minicomputer'' in
1973 [\H].  The Multum produced an answer after more than an hour, but
she remained uncertain whether other solutions were possible.  Now,
with the dancing links approach described above, we can obtain several
solutions almost instantly, and the total number of solutions turns
out to be~212.  The solutions fall into four classes, depending on
the behavior at the four corners; representatives of each achievable
class are shown in Figure~6.

\goodbreak
\newsection Applications to hexiamonds.
In the late 1950s, T.~H. O'Beirne introduced a pleasant variation on
polyominoes by substituting triangles for squares.  He named the
resulting shapes {\it polyiamonds\/}: moniamonds, diamonds, triamonds,
tetriamonds, pentiamonds, hexiamonds, etc.  The twelve hexiamonds were
independently discovered by J.~E. Reeve and J.~A. Tyrell [\RT], who
found more than forty ways to arrange them into a $6\times6$ rhombus.
Figure~7 shows one such arrangement, together with some arrow
dissections that I couldn't resist trying when I first learned about
hexiamonds.  The $6\times6$ rhombus can be tiled by the twelve
hexiamonds in exactly 156~ways.  (This fact was first proved by P.~J.
Torbijn [\T], who worked without a computer; \algoDLX\ confirms his
result after making 37{,}313{,}405 updates, if we restrict the
``sphinx'' to only 3 of its 12~orientations.)

\midinsert
\def\\#1#2#3#4{
\vbox{\halign{\hfil##\hfil\cr
 \epsfbox{\dancefig.7#1}\cr
 \noalign{\smallskip}
 \labels #2 solutions, \ #3 nodes, \ #4 updates\cr}}}

\line{\\3{4}{6,677}{4,687,159}\hfil\\2{0}{7,603}{3,115,387}}
\vskip22pt
\centerline{\\1{156}{70,505}{37,313,405}\hskip1.5in}
\vskip-22pt
\line{\\5{41}{35,332}{14,948,759}\hfil\\4{3}{5546}{3,604,817}}
\medskip
\centerline{\caption Figure 7. The twelve hexiamonds, packed into}
\centerline{a rhombus and into various arrowlike shapes.}
\endinsert

O'Beirne was particularly fascinated by the fact that seven of the
twelve hexiamonds have different shapes when they are flipped over,
and that the resulting 19 {\it one-sided hexiamonds\/} have the correct
number of triangles to form a hexagon: a hexagon of hexiamonds (see
Figure~8).  In November of 1959, after three months of trials, he
found a solution; and two years later he challenged the readers of
{\sl New Scientist\/} to match this feat [\OBi, \OBii, \OBiii].

Meanwhile he had shown the puzzle to Richard Guy and his family. The
Guys published several solutions in a journal published in Singapore,
where Richard was a professor [\Gi].  Guy, who has told the story
of this fascinating recreation in [\Gii], says that when O'Beirne
first described the puzzle, ``Everyone wanted to try it at once.  No
one went to bed for about 48~hours.''

\topinsert
\def\\#1#2#3#4#5#6#7{
\vbox{\halign{\hfil##\hfil\cr
 (#7)\cr
 \noalign{\vskip2pt}
 \epsfxsize=1.6in
 \epsfbox{\dancefig.8#1}\cr
 \noalign{\smallskip}
 \labels (hsym = #2, vsym = #3)\cr
 \labels #4 solutions, #5 nodes\cr
 \labels #6 updates\cr}}}

\line{\hfil\\0{51}{24}{1,914}{4,239,132}{2,142,276,414}a\hfil
           \\1{52}{24}{5,727}{21,583,173}{11,020,236,507}b\hfil}
\medskip
\line{\\2{32}{50}{11,447}{20,737,702}{10,315,775,812}c\hfil
      \\3{51}{22}{7,549}{24,597,239}{12,639,698,345}d\hfil
      \\4{48}{30}{6,675}{17,277,362}{8,976,245,858}e}
\medskip
\line{\hfil\\5{52}{27}{15,717}{43,265,607}{21,607,912,011}f\hfil
      \\6{48}{29}{75,490}{137,594,347}{67,723,623,547}g\hfil}
\medskip
\centerline{\caption Figure 8. Solutions to O'Beirne's hexiamond hexagon problem,}
\centerline{with the small hexagon at various distances from the center
            of the large one.}
\endinsert

A 19-level backtrack tree with many possibilities at each level makes
an excellent test case for the dancing links approach to covering, so
I fed O'Beirne's problem to my program.  I broke the general case into
seven subcases, depending on the distance of the hexagon piece from
the center; furthermore, when that distance was zero, I considered two
subcases depending on the position of the ``crown.''  Figure~8 shows a
representative of each of the seven cases, together with statistics
about the search.  The total number of updates performed was
134{,}425{,}768{,}494.

My goal was not only to count the solutions, but also to find
arrangements that were as symmetrical as possible---in response to a
problem that was stated in Berlekamp, Guy, and Conway's book {\sl
Winning Ways\/} [\BGC, page 788].  Let us define the {\it horizontal
sym\-metry\/} of a configuration to be the number of edges between pieces
that also are edges between pieces in the left-right reflection
of that configuration.  The overall hexagon has 156~internal edges,
and the 19~one-sided hexiamonds have 96~internal non-edges.  Therefore
if an arrangement were perfectly symmetrical---unchanged by left-right
reflection---its horizontal symmetry would be~60.  But no such
perfectly symmetric solution is possible.

The vertical symmetry of a
configuration is defined similarly, but with respect to top-bottom
reflection.  A solution to the hexiamond problem is {\it maximally
symmetric\/} if it has the highest horizontal or vertical symmetry
score, and if the smaller score is as large as possible consistent
with the larger score.  Each of the solutions shown in Figure~8 is, in
fact, maximally symmetric in its class.  (And so is the solution to
Dana Scott's problem that is shown in Figure~1: It has vertical
symmetry~36 and horizontal symmetry~30.)

The largest possible vertical symmetry score is~50; it is achieved in
Figure~8(c), and in seven other solutions obtained by independently
rearranging three of its symmetrical subparts.  Four of the eight have
a horizontal symmetry score of~32; the others have horizontal
symmetry~24. John Conway found these solutions by hand in 1964 and
conjectured that they were maximally symmetric overall.  But that
honor belongs uniquely to the solution in Figure~8(f), at least by my
definition, because Figure~8(f) has horizontal symmetry~52 and
vertical symmetry~27.  The only other ways to achieve horizontal
symmetry~52 have vertical symmetry scores of 20, 22, and 24. (Two of
those other ways do, however, have the surprising property that 13 of
their 19 pieces are unchanged by horizontal reflection; this is
symmetry of entire pieces, not just of edges.)

After I had done this enumeration, I read Guy's paper [\Gii] for the first time
and learned that Marc~M. Paulhus had already enumerated all solutions
in May 1996 [\P].  Good, our independent computations would confirm
the results.  But no---my program found 124{,}519 solutions, while his
had found 124{,}518! He reran his program in 1999 and now we agree.
\looseness=-1

O'Beirne [\OBii] also suggested an analogous problem for pentominoes,
since there are 18~one-sided pentominoes.  He asked if they can be put
into a $9\times10$ rectangle, and Golomb provided an example in [\G,
Chapter~6].  Jenifer Leech wrote a program to prove that there are
exactly 46~different ways to pack the one-sided pentominoes in a
$3\times30$ rectangle; see~[\M].  Figure~9 shows a maximally symmetric
example (which isn't really very symmetrical).

\midinsert
\epsfxsize=\hsize \epsfbox{\dancefig.9}
\centerline{\labels 46 solutions, \ 605,440 nodes, \ 190,311,749 updates,
\ hsym = 51, \ vsym = 48}
\medskip
\centerline{\caption Figure 9. The one-sided pentominoes, packed into a $3\times30$
  rectangle.}
\endinsert

I set out to count the solutions to the $9\times10$, figuring that an
18-stage exact cover problem with six 1s per row would be simpler than
a 19-stage problem with seven 1s per row.  But I soon found that the
task would be hopeless, unless I invented a much better algorithm.  The
Monte Carlo estimation procedure of [\MC] suggests that about
19~quadrillion updates will be needed, with 64~trillion nodes in the
search trees.  If that estimate is correct, I could have the result in
a few months; but I'd rather try for a new Mersenne prime.

I do, however, have a conjecture about the solution that will have
maximum horizontal symmetry; see Figure~10.

\midinsert
\centerline{\epsfbox{\dancefig.10}}
\centerline{\labels hsym = 74, \ vsym = 49}
\medskip
\centerline{\caption Figure 10. Is this the most symmetrical way}
\centerline{to pack one-sided pentominoes into a rectangle?}
\endinsert

\newsection A failed experiment.
Special arguments based on ``coloring'' often give important insights
into tiling problems.  For example, it is well known [\MB, pages 142
and~394] that if we remove two cells from opposite corners of a
chessboard, there is no way to cover the remaining 62~cells with
dominoes.  The reason is that the mutilated chessboard has, say,
32~white cells and 30~black cells, but each individual domino covers
one cell of each color.  If we present such a covering problem to \algoDLX,
it makes 4{,}780{,}846 updates (and finds 13{,}922 ways to place 30 of
the 31 dominoes) before concluding that there is no solution.
\looseness=-1

The cells of the hexiamond-hexagon problem can be colored black and
white in a similar fashion: All triangles that point left are black,
say, and all that point right are white.  Then fifteen of the
one-sided hexiamonds cover three triangles of each color; but the
remaining four, namely the ``sphinx'' and the ``yacht'' and their
mirror images, each have a four-to-two color bias.  Therefore every
solution to the problem must put exactly two of those four pieces into
positions that favor black.

I thought I'd speed things up by dividing the problem into six
subproblems, one for each way to choose the two pieces that will favor
black.  Each of the subproblems was expected to have about $1/6$~as
many solutions as the overall problem, and each subproblem was simpler
because it gave four of the pieces only half as many options as
before.  Thus I expected the subproblems to run up to 16~times as fast
as the original problem, and I expected the extra information about
impossible correlations of piece placement to help \algoDLX\ make
intelligent choices.

But this turned out to be a case where mathematics gave me bad advice.
The overall problem had 6675 solutions and required
8{,}976{,}245{,}858 updates (Figure~8(c)).  The six subproblems turned
out to have respectively 955, 1208, 1164, 1106, 1272, and 970
solutions, roughly as expected; but they each required between 1.7 and
2.2~billion updates, and the total work to solve all six subproblems
was 11{,}519{,}571{,}784.  So much for {\it that\/} bright idea.

\newsection Applications to tetrasticks.
Instead of making pieces by joining squares or triangles together,
Brian Barwell [\B] considered making them from line segments or sticks.  He
called the resulting objects {\it polysticks}, and noted that there
are 2~disticks, 5~tristicks, and 16~tetrasticks.  The tetrasticks are
especially interesting from a recreational standpoint; I received an
attractive puzzle in 1993 that was equivalent to placing ten of the
tetrasticks in a $4\times4$ square [\TP], and I spent many hours
trying to psych it out.

Barwell proved that the sixteen tetrasticks cannot be assembled into
any symmetrical shape.  But by leaving out any one of the five
tetrasticks that have an excess of horizontal or vertical line
segments, he found ways to fill a $5\times5$ square.  (See Figure~11.)
Such puzzles are quite difficult to do by hand, and he had found only
five solutions at the time he wrote his paper; he conjectured that
fewer than a hundred solutions would actually exist.
(The set of all solutions was first found by Wiezorke and
Haubrich~[\WH], who invented the puzzle independently
after seeing~[\TP].)

\midinsert
\def\\#1#2#3#4#5{
\vbox{\halign{\hfil##\hfil\cr
 \hfil(#5) \cr
 \noalign{\vskip2pt}
 \epsfxsize=.27\hsize \epsfbox{\dancefig.110#1}\cr
 \noalign{\smallskip}
 \labels #2 solutions, \ #3 nodes\cr
 \labels #4 updates\cr}}}
\line{\hfil\\1{72}{1,132,070}{283,814,227}a\hfil
           \\2{382}{3,422,455}{783,928,340}b\hfil}
\medskip
\line{\\3{607}{2,681,188}{611,043,121}c\hfil
      \\4{530}{3,304,039}{760,578,623}d\hfil
      \\5{204}{1,779,356}{425,625,417}e}
\medskip
\centerline{\caption Figure 11. Filling a $5\times5$ grid with 15 of the 16 tetrasticks;}
\centerline{we must leave out either the H, the J, the L, the N, or the Y.}
\endinsert

Polysticks introduce a new feature that is not present in the
polyomino and polyiamond problems: {\it The pieces must not cross each
other.}  For example, Figure~12 shows a non-solution to the problem
considered in Figure~11(c).
Every line segment in the grid of $5\times5$ squares is covered, but
the `V'~tetrastick crosses the~`Z'.

We can handle this extra complication by generalizing the exact cover
problem.  Instead of requiring all columns of a given 0\hair-1~matrix to be
covered by disjoint rows, we will distinguish two kinds of columns:
{\it primary\/} and {\it secondary}.  The generalized problem asks for a
set of rows that covers every primary column exactly once and
every secondary column {\it at most\/} once.

The tetrastick problem of Figure~11(c) can be set up as a generalized
cover problem in a natural way. First we introduce primary columns~F, H, I, J, N, O,
P, R, S, U, V, W, X, Y,~Z representing the fifteen tetrasticks
(excluding~${\rm L}$), as well as columns~${\rm H}xy$ representing the
\def\adj{\mathrel-\joinrel\joinrel\mathrel-} 
horizontal segments $(x,y)\adj(x+1,y)$ and ${\rm V}xy$ representing the
vertical segments $(x,y)\adj(x,y+1)$, for $0\le x,y<5$.  We also need
secondary columns ${\rm I}\hair xy$ to represent interior junction points
$(x,y)$, for $0<x,y<5$.  Each row represents a possible placement of a
piece, as in the polyomino and polyiamond problems; but if a piece has
two consecutive horizontal or vertical segments and does not lie on
the edge of the diagram, it should include the corresponding interior
junction point as well.

\midinsert
\epsfxsize=.195\hsize
\line{\hfil\epsfbox{\dancefig.12}\hfil
  \vbox{\hsize=.5\hsize \raggedright\noindent
        \caption Figure 12. Polysticks are not supposed to cross each other
        as they do here.\par\vskip24pt}\hfil}
\endinsert

For example, the two rows corresponding to the placement of ${\rm V}$
and ${\rm Z}$ in Figure~12 are
$$\matrix{{\rm V}&{\rm H}23&{\rm I}33&{\rm H}33&{\rm V}43&{\rm I}44&{\rm V}44\cr
          {\rm Z}&{\rm H}24&{\rm V}33&{\rm I}33&{\rm V}32&{\rm H}32\cr}$$
The common interior point~${\rm I}33$ means that these rows cross each
other.  On the other hand, ${\rm I}33$ is not a primary column,
because we do not necessarily need to cover it.  The solution in
Figure~11(c) covers only the interior points ${\rm I}14$,
${\rm I}21$, ${\rm I}32$, and ${\rm I}41$.

Fortunately, we can solve the generalized cover problem by using
almost the same algorithm as before.  The only difference is that we
initialize the data structure by making a circular list of the column
headers for the primary columns only.  The header for each secondary
column should have $L$ and~$R$ fields that simply point to
itself.  The remainder of the algorithm proceeds exactly as before, so
we will still call it \algoDLX.

A generalized cover problem can be converted to an equivalent exact
cover problem if we simply append one row for each secondary column,
containing a single~1 in that column. But we are better off working
with the generalized problem, because the generalized algorithm is
simpler and faster.

I decided to experiment with the subset of {\it welded tetrasticks},
namely those that do not form a simple connected path
because they contain junction points: ${\rm F, H, R, T, X, Y}$.  There
are ten {\it one-sided\/} welded tetrasticks if we add the mirror images
of the unsymmetrical pieces as we did for one-sided hexiamonds and
pentominoes.  And---aha---these ten tetrasticks can be arranged in a
$4\times4$ grid. (See Figure~13.)  Only three solutions are possible,
including the two perfectly symmetric solutions shown.  I've decided
not to show the third solution, which has the ${\rm X}$~piece in the
middle, because I want readers to have the pleasure of finding it for
themselves.

\midinsert
\line{\hfil\epsfxsize=.23\hsize\epsfbox{\dancefig.131}\hfil
   \epsfxsize=.23\hsize\epsfbox{\dancefig.132}\hfil}
\medskip
\centerline{\caption Figure 13. Two of the three ways to pack the}
\centerline{one-sided welded tetrasticks into a square.}
\endinsert

There are fifteen one-sided {\it unwelded tetrasticks}, and I thought
they would surely fit into a $5\times5$ grid in a similar way; but
this turned out to be impossible. The reason is that if, say,
piece~$\rm I$ is placed vertically, four of the six pieces $\rm J$,
$\rm J'$, $\rm L$, $\rm L'$, $\rm N$, $\rm N'$ must be placed to favor
the horizontal direction, and this severely limits the
possibilities. In fact, I have been unable to pack those fifteen
pieces into any simple symmetrical shape; my best effort so far is the
``oboe'' shown in Figure~14.

\midinsert
\epsfxsize=.9\hsize
\centerline{\epsfbox{\dancefig.14}}
\centerline{\caption Figure 14. The fifteen one-sided unwelded tetrasticks.}
\endinsert

I also tried unsuccessfully to pack all 25 of the one-sided
tetrasticks into the Aztec diamond pattern of Figure~15; but
I see no way to prove that a solution is impossible.
An exhaustive search seems out of the question at the present time.

\topinsert
\epsfxsize=.2\hsize
\line{\hfil\epsfbox{\dancefig.15}\hfil
  \vbox{\hsize=.5\hsize \raggedright\noindent
          \caption Figure 15. Do all 25 one-sided tetrasticks fit
           in this shape?\par\vskip24pt}\hfil}
\endinsert

\newsection Applications to queens.
Now we can return to the problem that led Hitotumatu and Noshita to
introduce dancing links in the first place, namely the $N$~queens
problem, because that problem is actually a special case of the
generalized cover problem in the previous section.  For example, the
4~queens problem is just the task of covering eight primary columns $({\rm
R}0, {\rm R}1, {\rm R}2, {\rm R}3, {\rm F}0, {\rm F}1, {\rm F}2, {\rm
F}3)$ corresponding to ranks and files,
while using at most one element in each of the secondary columns
$({\rm A}0, {\rm A}1, {\rm A}2, {\rm A}3, {\rm A}4, {\rm A}5, {\rm
A}6,\allowbreak {\rm B}0, {\rm B}1, {\rm B}2, {\rm B}3, {\rm B}4, {\rm B}5, {\rm
B}6)$ corresponding to diagonals, given the sixteen rows
$$\matrix{{\rm R}0&{\rm F}0&{\rm A}0&{\rm B}3\cr
          {\rm R}0&{\rm F}1&{\rm A}1&{\rm B}4\cr
          {\rm R}0&{\rm F}2&{\rm A}2&{\rm B}5\cr
          {\rm R}0&{\rm F}3&{\rm A}3&{\rm B}6\cr
          {\rm R}1&{\rm F}0&{\rm A}1&{\rm B}2\cr
          {\rm R}1&{\rm F}1&{\rm A}2&{\rm B}3\cr
          {\rm R}1&{\rm F}2&{\rm A}3&{\rm B}4\cr
          {\rm R}1&{\rm F}3&{\rm A}4&{\rm B}5\cr
          {\rm R}2&{\rm F}0&{\rm A}2&{\rm B}1\cr
          {\rm R}2&{\rm F}1&{\rm A}3&{\rm B}2\cr
          {\rm R}2&{\rm F}2&{\rm A}4&{\rm B}3\cr
          {\rm R}2&{\rm F}3&{\rm A}5&{\rm B}4\cr
          {\rm R}3&{\rm F}0&{\rm A}3&{\rm B}0\cr
          {\rm R}3&{\rm F}1&{\rm A}4&{\rm B}1\cr
          {\rm R}3&{\rm F}2&{\rm A}5&{\rm B}2\cr
          {\rm R}3&{\rm F}3&{\rm A}6&{\rm B}3\rlap{\thinspace.}\cr}$$
In general, the rows of the 0\hair-1 matrix for the $N$~queens problem are 
$$\advance\abovedisplayskip-6pt 
  \advance\belowdisplayskip-6pt 
{\rm R}i\quad {\rm F}j\quad {\rm A}(i+j)\quad {\rm B}(N-1-i+j)$$
for $0\le i, j<N$.  (Here ${\rm R}i$ and ${\rm F}j$ represent ranks
and files of a chessboard; ${\rm A}k$ and ${\rm B}\ell$ represent
diagonals and reverse diagonals.  The secondary columns ${\rm A}0,
{\rm A}(2N-2), {\rm B}0$, and ${\rm B}(2N-2)$ each arise in only one
row of the matrix so they can be omitted.)

When we apply \algoDLX\ to this generalized cover problem, it behaves
quite differently from the traditional algorithms for the $N$~queens
problem, because it branches
sometimes on different ways to occupy a rank of the chessboard and
sometimes on different ways to occupy a file.  Furthermore, we gain
efficiency by paying attention to the order in which primary columns
of the cover problem are considered when those columns all have the
same $S$~value (the same branching factor): It is better to place
queens near the middle of the board first, because central positions
rule out more possibilities for later placements.

Consider, for example, the eight queens problem.  Figure~16(a) shows
an empty board, with 8~possible ways to occupy each rank and each
file.  Suppose we decide to place a queen in~${\rm R}4$ and ${\rm
F}7$, as shown in Figure~16(b).  Then there are five ways to
cover~${\rm F}4$; after choosing ${\rm R}5$ and ${\rm F}4$,
Figure~16(c), there are four ways to cover~${\rm R}3$, and so on.  At
each stage we choose the most constrained rank or file, using the
``organ pipe ordering''
$$\advance\abovedisplayskip-6pt 
  \advance\belowdisplayskip-6pt 
  {\rm R}4\ {\rm F}4\ {\rm R}3\ {\rm F}3\ {\rm R}5\ {\rm F}5\ {\rm R}2\ {\rm F}2\ {\rm R}6\ {\rm F}6\ {\rm R}1\ {\rm F}1\ {\rm R}7\ {\rm F}7\ {\rm R}0\ {\rm F}0$$
to break ties.  Placing a queen in ${\rm R}2$ and ${\rm F}3$ after
Figure~16(d) makes it impossible to cover~{\rm F}2, so backtracking
will occur even though only four queens have been tentatively placed.

\midinsert
\def\\#1#2{\vtop{\halign{\hfil##\hfil\cr
 (#2)\cr
 \noalign{\smallskip}
 \def\epsfsize##1##2{.8##1}\epsfbox{\dancefig.16#1}\cr
}}}

\line{\hfil\\0a\hfil\\1b\hfil}
\medskip
\line{\hfil\\2c\hfil\\3d\hfil}
\medskip
\centerline{\caption Figure 16. Solving the 8 queens problem by treating
  ranks and files symmetrically.}
\endinsert

The order in which header nodes are linked together at the start of
\algoDLX\ can have a significant effect on the running time.  For
example, experiments on the 16~queens problem show that the search
tree has 312{,}512{,}659 nodes and requires 5{,}801{,}583{,}789
updates, if the ordering ${\rm R}0\ {\rm R}1\ \ldots\ {\rm R}15\ {\rm
F}0\ {\rm F}1\ \ldots\ {\rm F}15$ is used, while the organ-pipe
ordering ${\rm R}8\ {\rm F}8\ {\rm R}7\ {\rm F}7\ {\rm R}9\ {\rm F}9\
\ldots\ {\rm R}0\ {\rm F}0$ requires only about 54\% as many updates.
On the other hand, the order in which individual elements of a row or
column are linked together has no effect on the algorithm's total
running time.

Here are some statistics observed when \algoDLX\ solved small cases of
the $N$~queens problem using organ-pipe order, without reducing the
number of solutions by taking symmetries of the board into account:
$$\vbox{\halign{\hfil#\quad&\hfil#\quad&\hfil#\quad&\hfil#\quad&\hfil#\quad&\hfil#\cr
$N$&Solutions\enspace&Nodes\enspace&Updates\enspace
&R-Nodes\enspace&R-Updates\enspace\cr
1&           1&             2&               3&              2&               3\cr
2&           0&             3&              19&              3&              19\cr
3&           0&             4&              56&              6&              70\cr
4&           2&            13&             183&             15&             207\cr
5&          10&            46&             572&             50&             626\cr
6&           4&            93&           1{,}497&            115&           1{,}765\cr
7&          40&           334&           5{,}066&            376&           5{,}516\cr
8&          92&         1{,}049&          16{,}680&          1{,}223&          18{,}849\cr
9&         352&         3{,}440&          54{,}818&          4{,}640&          71{,}746\cr
10&         724&        11{,}578&         198{,}264&         16{,}471&         269{,}605\cr
11&       2{,}680&        45{,}393&         783{,}140&         67{,}706&       1{,}123{,}572\cr
12&      14{,}200&       211{,}716&       3{,}594{,}752&        312{,}729&       5{,}173{,}071\cr
13&      73{,}712&     1{,}046{,}319&      17{,}463{,}157&      1{,}589{,}968&      26{,}071{,}148\cr
14&     365{,}596&     5{,}474{,}542&      91{,}497{,}926&      8{,}497{,}727&     139{,}174{,}307\cr
15&   2{,}279{,}184&    31{,}214{,}675&     513{,}013{,}152&     49{,}404{,}260&     800{,}756{,}888\cr
16&  14{,}772{,}512&   193{,}032{,}021&   3{,}134{,}588{,}055&    308{,}130{,}093&   4{,}952{,}973{,}201\cr
17&  95{,}815{,}104& 1{,}242{,}589{,}512&  20{,}010{,}116{,}070&  2{,}015{,}702{,}907&  32{,}248{,}234{,}866\cr 
18& 666{,}090{,}624& 8{,}567{,}992{,}237& 141{,}356{,}060{,}389& 13{,}955{,}353{,}609& 221{,}993{,}811{,}321\cr}}$$
Here ``R-nodes'' and ``R-Updates'' refer to the results
when we consider only ${\rm R}0$, ${\rm R}1$, \dots, ${\rm R}(N-1)$ to be
primary columns that need to be covered; columns ${\rm F}j$ are
secondary. In this case the algorithm
reduces to the usual procedure in which branching occurs only on ranks
of the chessboard.  The advantage of mixing rows with columns becomes
evident as $N$~increases, but I'm not sure whether the ratio of
R-Updates to Updates will be unbounded or approach a
limit as $N$~goes to infinity.

I should point out that special methods are known for counting the number of
solutions to the $N$~queens problem without actually generating the
queen placements [\RVZ].

\newsection Concluding remarks.
\AlgoDLX, which uses dancing links to implement the ``natural''
algorithm for exact cover problems, is an effective way to enumerate
all solutions to such problems.  On small cases it is nearly as fast
as algorithms that have been tuned to solve particular classes of
problems, like pentomino packing or the $N$~queens problem, where
geometric structure can be exploited.  On large cases it appears to
run even faster than those special-purpose algorithms, because of its
ordering heuristic.  And as computers get faster and faster, we are of
course tackling larger and larger cases all the time.

In this paper I have used the exact cover problem to illustrate the
versatility of dancing links, but I could have chosen many other
backtrack applications in which the same ideas apply.  For example,
the approach works nicely with the Waltz filtering algorithm [\W];
perhaps this fact has subliminally influenced my choice of names.  I
recently used dancing links together with a dictionary of about 600
common three-letter words of English to find word squares such as
$$\def\\#1#2#3#4#5#6#7#8#9{\vcenter{\halign{&\tt\kern.1em##\kern.1em\cr
 #1&#2&#3\cr#4&#5&#6\cr#7&#8&#9\cr}}}
\\ATEWINLED \qquad
\\BEDOARWRY \qquad
\\OHMRUEBET \qquad
\\PEAURNBAY \qquad
\\TWOIONTEE$$
in which each row, column, and diagonal is a word; about 60 million
updates produced all solutions.
I~believe that a terpsichorean technique is significantly better than
the alternative of copying the current state at every level, as
considered in the pioneering paper by Haralick and Elliott on
constraint satisfaction problems [\HE].  Certainly the use of (1) and~(2)
is simple, useful, and fun.

\medskip
``What a dance / do they do / Lordy, I am tellin' you!'' [\MM]

\newsection Acknowledgments.
I wish to thank Sol Golomb, Richard Guy, and Gene Freuder for the help
they generously gave me as I was preparing this paper.
Maggie McLoughlin did an excellent job of translating my scrawled
manuscript into a well-organized \TeX\ document. And I
profoundly thank Tomas Rokicki, who provided the new computer on which
I did most of the experiments, and on which I hope to keep links
dancing merrily for many years.

\newsection Historical notes.
(1)~Although the IAS computer was popularly known in Princeton as the
``{\mc MANIAC},'' that title properly belonged only to a similar but
different series of computers built at Los Alamos.  (See [\MW].) \
(2)~George Jelliss [\J] discovered that the great puzzle masters H.~D.
Benjamin and T.~R. Dawson experimented with the concept of polysticks
already in 1946--1948.  However, they apparently did not publish any
of their work. \
(3)~My names for the tetrasticks are slightly
different from those originally proposed by Barwell~[\B]: I prefer to
use the letters ${\rm J,\ R,}$ and ${\rm U}$ for the pieces he called
$\rm U$, $\rm J$, and~${\rm C}$, respectively.

\newsection Program notes.
The implementation of \algoDLX\ that I used when preparing this
paper is file {\tt dance.w} on webpage
{\tt http:/\kern-.1em/www-cs-faculty.stanford.edu/\char`\~knuth/\allowbreak
programs.html}.
See also the related files {\tt polyominoes.w}, {\tt polyiamonds.w},
{\tt polysticks.w}, and {\tt queens.w}.

\vfill\eject

\centerline{\bf References}
\medskip
\def\ref[#1] {\smallskip
        \noindent\hbox to\parindent{\hss[#1]\enspace}%
           \hangindent=\parindent
           \hangafter=1\relax}

\ref[\TP]  {\sl 845 Combinations Puzzles:  845 Interestingly
Combinations\/} (Taiwan:\ R.O.C. Patent 66009). [There is no indication of
the author or manufacturer. This puzzle, which is available from
{\tt www.puzzletts.com}, actually has only 83 solutions. It carries a
Chinese title, ``Dr.~Dragon's Intelligence Profit System.'']

\ref[\MM] Harry Barris, {\sl Mississippi Mud\/} (New York:\ Shapiro,
Bernstein \& Co., 1927).

\ref[\B] Brian R. Barwell, ``Polysticks,'' {\sl Journal of
Recreational Mathematics\/ \bf22} (1990), 165--175.

\ref[\BGC] Elwyn R. Berlekamp, John H. Conway, and Richard K. Guy,
{\sl Winning Ways for Your Mathematical Plays\/ \bf2} (London:\
Academic Press, 1982).

\ref[\MB] Max Black, {\sl Critical Thinking\/} (Englewood Cliffs, New
Jersey:\ Prentice--Hall, 1946).  [Does anybody know of an earlier
reference for the problem of the ``mutilated chessboard''?]

\ref[\DDH] Ole-Johan Dahl, Edsger W. Dijkstra, and C.~A.~R. Hoare, {\sl
Structured Programming\/} (London:\ Academic Press, 1972).

\ref[\deBe] N.~G. de Bruijn, personal communication (9 September 1999):
``\dots\ it was almost my first activity in programming that I got all 2339
solutions of the $6\times10$ pentomino on an IBM1620 in March 1963
in 18 hours. It had to cope with the limited memory of that machine,
and there was not the slightest possibility to store the full matrix \dots\
But I could speed the matter up by having a very long program, and that one
was generated by means of another program.''

\ref[\deB] N.~G. de Bruijn, ``Programmeren van de pentomino puzzle,''
{\sl Euclides\/ \bf47} (1971/72), 90--104.

\ref[\D] Henry Ernest Dudeney, ``74.---The broken chessboard,'' in {\sl
The Canterbury Puzzles\/}, (London:\ William Heinemann, 1907), 90--92,
174--175.

\ref[\Fle] John G. Fletcher, ``A program to solve the pentomino
problem by the recursive use of macros,'' {\sl Communications of the
ACM\/ \bf8} (1965), 621--623.

\ref[\Flo] Robert W. Floyd, ``Nondeterministic algorithms,'' {\sl
Journal of the ACM\/ \bf14} (1967), 636--644.

\ref[\MG] Martin Gardner, ``Mathematical games: More about complex
dominoes, plus the answers to last month's puzzles,'' {\sl Scientific
American\/ \bf197},\thinspace6 (December 1957), 126--140.

\ref[\GJ] Michael R.\ Garey and David S.\ Johnson, {\sl Computers and
Intractability\/} (San Francisco:\ Freeman, 1979).

\ref[\GO] Solomon W. Golomb, ``Checkerboards and polyominoes,'' {\sl
American Mathematical Monthly\/ \bf 61} (1954), 675--682.

\ref[\G] Solomon W.\ Golomb, {\sl Polyominoes}, second edition
(Princeton, New Jersey:\  Princeton University Press, 1994).

\ref[\GB] Solomon W.\ Golomb and Leonard D. Baumart, ``Backtrack
programming,'' {\sl Journal of the ACM\/ \bf12} (1965), 516--524.

\ref[\Gi] Richard K. Guy, ``Some mathematical recreations,'' {\sl
Nabla\/} (Bulletin of the Malayan Mathematical Society) {\bf7} (1960),
97--106, 144--153.

\ref[\Gii] Richard K. Guy, ``O'Beirne's Hexiamond,'' in {\sl The
Mathemagician and Pied Puzzler\/}, edited by Elwyn Berlekamp and Tom
Rodgers (Natick, Massachusetts:\ A.~K. Peters, 1999), 85--96.

\ref[\HE] Robert M. Haralick and Gordon L. Elliott, ``Increasing tree
search efficiency for constraint satisfaction problems,'' {\sl
Artificial Intelligence\/ \bf14} (1980), 263--313.

\ref[\H] Jenifer Haselgrove, ``Packing a square with Y-pentominoes,''
{\sl Journal of Recreational Mathematics\/ \bf7} (1974), 229.

\ref[\HH] C.~B. and Jenifer Haselgrove, ``A computer program for
pentominoes,'' {\sl Eureka\/ \bf23},\thinspace2 (Cambridge, England:\ The
Archimedeans, October 1960), 16--18.

\ref[\HN] Hirosi Hitotumatu and Kohei Noshita, ``A technique for
implementing backtrack algorithms and its application,'' {\sl
Information Processing Letters\/ \bf 8} (1979), 174--175.

\ref[\J] George P.\ Jelliss, ``Unwelded polysticks,'' {\sl Journal of
Recreational Mathematics\/ \bf29} (1998), 140--142.

\ref[\MC] Donald E. Knuth, ``Estimating the efficiency of backtrack
programs,'' {\sl Mathematics of Computation\/ \bf29} (1975), 121--136.

\ref[\TTP] Donald E. Knuth, {\sl \TeX: The Program\/} (Reading,
Massachusetts:\ Addison--Wesley, 1986).

\ref[\M] Jean Meeus, ``Some polyomino and polyamond 
problems,'' {\sl Journal of Recreational Mathematics\/ \bf6} (1973), 215--220.

\ref[\MW] N. Metropolis and J. Worlton, ``A trilogy of errors in the
history of computing,'' {\sl Annals of the History of Computing\/ \bf2}
(1980), 49--59.

\ref[\OBi] T.~H. O'Beirne, ``Puzzles and Paradoxes 43: Pell's equation
in two popular problems,'' {\sl New Scientist\/ \bf12} (1961),
260--261.

\ref[\OBii] T.~H. O'Beirne, ``Puzzles and Paradoxes 44: Pentominoes
and hexiamonds,'' {\sl New Scientist\/ \bf12} (1961), 316--317.  [``So
far as I know, hexiamond has not yet been put through the mill on a
computer; but this could doubtless be done.'']

\ref[\OBiii] T.~H. O'Beirne, ``Puzzles and Paradoxes 45: Some
hexiamond solutions:  and an introduction to a set of 25 remarkable
points,'' {\sl New Scientist\/ \bf12} (1961), 379--380.

\ref[\P] Marc Paulhus, \kern-2pt``Hexiamond Homepage,'' 
{\tt http:/\kern-.15em/www\kern-1pt.math.\kern-.05emucalgary\kern-1pt.ca/\char`\~paulhusm/\break hexiamond1}.

\ref[\RT] J.~E. Reeve and J.~A. Tyrell, ``Maestro puzzles,'' {\sl The
Mathematical Gazette\/ \bf45} (1961), 97--99.

\ref[\RVZ] Igor Rivin, Ilan Vardi, and Paul Zimmermann, ``The
$n$-queens problem,'' {\sl American Mathematical Monthly\/ \bf101}
(1994), 629--639.

\ref[\S]  Dana S. Scott, \kern-1.5pt``Programming a combinatorial puzzle,''
Technical Report No.\thinspace1 (Princeton, New Jersey:\ Princeton University
Department of Electrical Engineering, 10 June 1958), $\rm ii+14+5$ pages.
[From page 10:  ``$\ldots$ the main problem in the program was to
handle several lists of indices that were continually being
modified.'']

\ref[\T] P.~J. Torbijn, ``Polyiamonds,'' {\sl Journal of Recreational
Mathematics\/ \bf2} (1969), 216--227.

\ref[\W]  David Waltz, ``Understanding line drawings of scenes with
shadows,'' in {\sl The Psychology of Computer Vision\/}, edited by
P.~Winston (New York:\ McGraw--Hill, 1975), 19--91.

\ref[\WH] Bernhard Wiezorke and Jacques Haubrich, ``Dr.\ Dragon's
polycons,'' {\sl Cubism For Fun\/ \bf33} (February 1994), 6--7.

\newsection Addendum.
During November, 1999,
Alfred Wassermann of Universit\"at Bayreuth succeeded in covering the
Aztec diamond of Figure~15
with one-sided tetrasticks, using a cluster of work\-stations
running \algoDLX. The 107 possible solutions, which are quite beautiful, have been
posted at {\tt http://did.mat.uni-bayreuth.de/wassermann/}.
He subsequently enumerated the 10,440,433 solutions to the $9\times10$ one-sided pentomino problem; many of these turn out to be more symmetric
than the one in Figure~10.

\bye